# Bluff body uses deep-reinforcement-learning trained active flow control to achieve hydrodynamic stealth

Feng Ren, Hui Tang[*]

Research Center for Fluid-Structure Interactions, Department of Mechanical Engineering, The Hong Kong Polytechnic University, Kowloon, Hong Kong, China

[*]Corresponding author: h.tang@polyu.edu.hk

**Abstract**

We propose a novel active-flow-control (AFC) strategy for bluff bodies to hide their hydrodynamic traces from predators. A group of windward-suction-leeward-blowing (WSLB) actuators are adopted to control the wake of a circular cylinder submerged in a uniform flow. An array of velocity sensors are deployed in the near wake to provide feedback signals. Through the data-driven deep reinforcement learning (DRL), effective control strategies are trained for the WSLB actuation to mitigate the cylinder's hydrodynamic signatures, i.e., strong shears and periodically shed vortices. Only a 0.29% deficit in streamwise velocity is detected, which is a 99.5% reduction from the uncontrolled value. The same control strategy is found to be also effective when the cylinder undergoes transverse vortex-induced vibration (VIV). The findings from this study can shed some lights on the design and operation of underwater structures and robotics to achieve hydrodynamic stealth.

Predator-prey interactions influence ecosystem structures. In aquatic communities, these interactions are strongly affected by disturbances in the surrounding liquid medium, especially when vision detection does not work well. One example is that harbor seals (*Phoca vitulina*) use whiskers to sense hydrodynamic perturbations produced by their preys, predators or conspecifics, thus helping to forage in dark or turbid water [1-3]. Another example is that fish use their lateral lines to perceive weak hydrodynamic perturbations induced by surrounding stationary or moving bodies [4]. Inspired by these aquatic animals, scientists and engineers have strived to develop sensitive sensors for underwater detection [5-7]. From the preys' perspective, on the other hand, if achieving hydrodynamic stealth of a certain level by mitigating or even eliminating their hydrodynamic traces, they can reduce the chance of being captured by predators. The hydrodynamic traces are usually in the form of shears or shed vortices, which are temporally unsteady and spatially uneven. In this letter, we propose a novel strategy, i.e., the use of machine-learning based AFC, to interfere the generation of these hydrodynamic traces, so that natural or artificial preys can achieve hydrodynamic stealth on need basis and escape from their predators.

*Problem description.* To demonstrate the strategy, we adopt a generic bluff body, i.e., a circular cylinder, for the control. As shown in Fig. 1(a), a circular cylinder with mass $m$ and diameter $D$ is immersed in a uniform flow of velocity $U$. The corresponding diameter-based Reynolds number is fixed at Re = 100, where asymmetric vortices will periodically shed from the cylinder, forming a Kármán vortex street. The cylinder's streamwise motion is refrained all the time while it is transversely connected to a spring of stiffness $K$. When $K$ is infinitely large, the cylinder will be stationary in the flow, whereas when $K$ is finite, the cylinder will undergo vortex-induced vibration (VIV) in the transverse direction governed by

$$m\ddot{y} = -Ky + F_L \tag{1}$$

where $F_L$ is the hydrodynamic lift due to the asymmetric vortex shedding. Two key dimensionless parameters, i.e., the mass ratio $m^*$ and the reduced velocity $U_R$, can be used to characterize the cylinder's dynamics

$$m^* = \frac{m}{\rho D^2}, \ \ U_R = \frac{U}{f_N D} \tag{2}$$

where $\rho$ is the fluid density and $f_N = (K/m)^{1/2}/2\pi$ is the natural frequency of this mass-spring system in vacuum. By varying $m^*$ and $U_R$, the cylinder may vibrate in a "lock-in" regime where the vibration amplitude is significantly large due to resonance, making the control very challenging [8]. Figs 1(b) to 1(d) show the uncontrolled wake patterns behind the stationary, locked-in vibrating, and unlocked-in vibrating cylinder, respectively, at the same instant, all revealing strong hydrodynamic signatures. Note that, due to the large-amplitude vibrations, the wake of the locked-in VIV cylinder seems quite different from the wakes in the other two cases.

To mitigate the cylinder's hydrodynamic traces, five pairs of WSLB actuators are adopted to provide control authority [9,10], as depicted in Fig. 1(a). In line with the streamwise direction, each pair of actuators consists of a suction slot at the windward side of the cylinder and a blowing slot at the leeward side. The five leeward actuators are located azimuthally at $\theta = \pm 80°$, $\pm 40°$, and $0°$, measured from the centerline. The actuator pair in the centerline has slots of $12°$ arc, while the other eight off-centerline actuators have a slot of $6°$ arc, altogether occupying one fifth of cylinder's surface. For each pair of actuators, fluid is horizontally sucked into the windward slot and horizontally injected out from the leeward slot with an identical velocity, hence realizing zero-net-mass-flux actuation. To simplify the control, the actuator pairs located at $\theta = \pm 80°$ are always operated in phase, so do the pairs located at $\theta = \pm 40°$. Hence only three independent controls are employed in the current study.

To provide feedback signals to the controller, an array of velocity sensors are placed $2D$ downstream of the cylinder, sensing the cylinder's hydrodynamic signatures through measuring the streamwise and transverse velocities at 33 equal-spacing locations along a vertical line of $2D$ length, as depicted in Fig. 1(a). This sensor array is chosen according to an experiment setup for drag measurement [11]. In practice, it can be a rake of hot-wire probes or Pitot tubes.

*Method.* The above control problem is studied in a numerical simulation environment, in which the flow is simulated by solving the incompressible Navier-Stokes equations. To achieve both high fidelity and computational efficiency, an in-house GPU-accelerated lattice Boltzmann solver is used. The details of this numerical solver can be found in Supplemental Material. Using an NVIDIA Tesla K40c GPU server, the averaged computational time for each trial control case can be managed within 6 minutes.

With the velocity information collected from the sensor array, a cost function $J$ is defined for the purpose of flow control

$$J = \sum_{i=1}^{N_S} \left[ \left( u_i - 1 \right)^2 + v_i^2 \right] + C \sum_{i=1}^{N_A} u_{a,i}^2 \qquad (3)$$

where $N_S$ and $N_A$ are the numbers of sensors and WSLB actuators, respectively. $u_i$ and $v_i$ are the instantaneous streamwise and transverse velocities read by the $i$th sensor, and $u_a$ is the actuation velocity. This cost function consists of two components. The first component, i.e., the first term of Eq. 3, is the velocity deficit, namely, the discrepancy in velocities between the wake and the uniform incoming flow (denoted as $J_1$ hereafter). $J_1 = 0$ when the wake is identical to the incoming flow, indicating a zero hydrodynamic trace. The second component reflects the energy consumed by the WSLB actuators for the control (the second term of Eq. 3), in which a small weighting coefficient $C = 10^{-4}$ is applied to prevent over fluctuation of the actuation. In addition, the actuation velocity $u_a$ is constrained within the range of [0, 5] to avoid overloading.

*DRL control.* Due to the unsteadiness, nonlinearity and high dimensionality of this flow control problem, it is challenging to find effective control laws in an explicit form. Instead, we adopt the DRL, a machine learning technique, to search feasible control strategies. The DRL uses artificial neural networks (ANNs) to approximate strongly nonlinear dynamics by learning through trial-and-error [12], which, besides the striking, high-profile victories in the game of Go against the best human players [13,14], has recently been successively applied in automatic control [15] and complex fluid-related problems, e.g., fish schooling [16], bird flying in turbulent environment [17], flow navigation of smart particles [18], etc. Reviews of recent successful DRL applications in fluid mechanics and AFC can be found in Rabault et al. [19] and Ren et al. [20].

In the present DRL framework, effective control strategies are learnt through interactions between the controller, i.e., an "agent", and the cylinder wake, i.e., the "fluid environment". As illustrated in Fig. 2, the agent feels the "state" of the environment using the sensor array, and exerts changes to the environment through "action", i.e., the WSLB actuations. To encourage the agent to keep improving its performance during the learning process, a "reward" is defined as $-J$, such that a better action gets a higher reward.

The proximal policy optimization (PPO) model [21], currently often regarded as the state-of-the-art algorithm for performing continuous control, is adopted to train the DRL agent through stochastic trial-and-error. In this model, two sets of ANNs are used in the agent, known as the "critic" and "actor" networks. The evolving "critic" network estimates the reward, whereas the evolving "actor" network models the agent's policy, i.e., the control strategy. Through this PPO-based DRL training, it is expected that the agent can learn effective control strategies in a similar way like humans. More details about the settings of this DRL framework can be found in Supplemental Material.

*Scenario I - stationary cylinder.* A successful learning process is shown in Fig. 3, in which the variation of time-averaged $J$ is presented against a number of episodes. Here an episode represents a complete simulation run with reinitialized flow environment and a control strategy updated according to the states, actions and rewards in previous episodes. It is seen that the control strategy improves in general with a fast rate in the first 500 episodes, and then gradually converges. Four representative episodes, i.e., Episode 1, 256, 493 and 1000, are chosen to show details of the WSLB actuation and the control effects during the learning process. In Episode 1, without any prior knowledge, the DRL agent manages to break down the asymmetric vortex shedding from the cylinder by operating the 0° (i.e., centerline) actuator with the strongest actuation and the ±80° actuators with the weakest. As a result, the originally unsteady wake becomes nearly steady, consisting of two strong shear layers mirrored about the cylinder's centerline, as shown in subplot (A) on the right. However, eminent velocity deficit in the wake can still be detected by the sensor array. As the learning proceeds and more experience is obtained, the DRL agent gradually reduces the 0° actuator's strength while increasing the ±40° actuators'. The two shear layers become less evident and the resulting velocity profile gradually approaches to a uniform one. Once a mature control strategy is learnt, the DRL agent decides to operate the ±40° actuators with the greatest strength, i.e., $u_{a,2} = u_{a,4} = 3.1$, and operate the 0° actuator with the smallest strength, i.e., $u_{a,3} = 2.3$. This strategy is somewhat consistent with what has been revealed in previous wake/VIV control studies: to make the control effective and efficient, the jet-like actuation should be applied close to the mean separation points, i.e., $\theta = \pm 58°$ at Re = 100 [10,22-24]. With this control strategy, the shear layers are no longer tangible and the measured velocity profile almost coincides with the uniform one as in the incoming flow (the mean discrepancy is only 0.29%, reduced by 99.5% if compared with the uncontrolled case), as shown in subplot (D) on the right in Fig. 3. A video showing the evolution of the flow field with this well-trained control strategy can be found in Supplementary Material. Through this control the cylinder's hydrodynamic traces are almost removed, and any predator just $2D$ downstream can hardly feel its existence. Furthermore, as revealed in the four insets, the fluctuations in the actuation are significantly mitigated when the learning process converges, indicating great confidence of the leant control strategy.

To further evaluate the performance of the learnt control strategy, temporal histories of the cost function, hydrodynamic forces and velocity fields obtained in Episode 1000 are presented in Fig. 4. It is seen from Fig. 4(a) that the control efficiently makes the cost function reach and maintain its lowest values just after about $t = 5$, slightly more than one oscillation cycle of the uncontrolled cylinder. The control also quickly reduces the hydrodynamic forces, in both streamwise and transverse directions, to values very close to zero, as shown in Fig. 4(b). The nearly zero transverse force, i.e., lift, reflects that the asymmetric vortex shedding has been successfully suppressed by the WSLB actuation. The slight oscillation around zero in the streamwise force reveals that the flow induced drag is almost balanced by the WSLB generated trust, indicating that the cylinder achieves the cruising status in the flow. This, from the momentum perspective, confirms the recovery of the uniform flow in the wake.

Figs. 4(c) and 4(d) compare the spatio-temporal distributions of the net streamwise velocity, $u - 1$, and transverse velocity, $v$, measured by the sensor array before and after the WSLB control. It is seen that, before the control, both the net streamwise and transverse velocities are spatio-temporally symmetric. Near the cylinder's centerline, the net streamwise velocity is always negative, reflecting the drag-production feature. After the control is switched on, the spatio-temporal patterns are quickly broken down, both evolving into nearly zero velocity fields.

*Scenario II - VIV cylinder.* The very success of the DRL selected control strategy for the stationary cylinder has motivated us to further explore the control in a more complicated scenario where the cylinder undergoes VIV in the transverse direction. In this scenario, the fixed sensor array has a motion relative to the cylinder, generating different readings as revealed in the velocity vectors shown in Figs. 1(b) to 1(d). This change may cause the deterioration of the WSLB-based control. Hence it is more challenging to learn a converged control strategy from the DRL training started with randomly initialized ANNs like what has been done in Scenario I, as evidenced in Fig. 4 of Supplemental Material. However, since in Scenario I the hydrodynamic lift is very close to zero, the same control strategy may be effective in suppressing the VIV first and then mitigating the hydrodynamic traces. Hence, the learnt control strategy in Scenario I, i.e., the well trained ANNs, is directly applied as the starting point in the new learning, lasting for 250 more episodes.

Results for the above transfer learning are presented in Fig. 5. It seems that the learnt control strategy in Scenario I already works well for Scenario II, as evident by the flat mean learning curves in the entire 250 episodes as shown in Fig. 5(a). With this type of control strategy, the cylinder's VIV can be suppressed to a large extent in both the locked-in and unlocked-in situations, as evidenced in Fig. 5(b) by the very small standard deviations in transverse displacement.

As expected, the control strategy obtained from transfer learning works well as long as the VIV is suppressed. It effectively mitigates the cylinder's hydrodynamic traces, reducing the mean velocity deficit by 96.5% in the locked-in case and 95.9% in the unlocked-in case, respectively, compared with the corresponding uncontrolled cases. The controlled wakes are shown in Figs. 5(c) and 5(d). Unlike in the controlled case in Scenario I, vortex shedding can still be observed in these two cases, mainly due to the imperfect VIV control. However, with the learnt control, this strong hydrodynamic signature is to a great extent constrained in the near wake, i.e., $x < 2D$, beyond which only very weak vorticities are observed. The remarkably weakened vorticities in the mid-wake will soon diffuse out and can be covered by background turbulence. Two videos showing the control effects for the VIV cylinder in the locked-in and unlocked-in situations can be found in Supplementary Material.

In summary, we demonstrated that using DRL trained AFC a bluff body can achieve hydrodynamic stealth from its predators. The well-trained WBLS control not only balances the body's hydrodynamic drag during cruising, but also greatly reduces the velocity deficit in the very near wake, hence mitigating or even eliminating the body's hydrodynamic traces. It was also shown that the trained WBLS control can effectively suppress VIV, and hence can also mitigate the hydrodynamic traces of VIV bodies. Although many parameters are underexplored, e.g., the effects of Reynolds number, WSLB actuator settings and sensor settings, the findings from this interesting study can shed some lights on the design and operation of underwater robotics for achieving low detectability.

**Acknowledgement**

This study was financially supported by Research Grants Council of Hong Kong under General Research Fund (Project No. 15249316 & 15214418), and the Departmental General Research Fund (Project No. G-YBXQ).

**Figures**

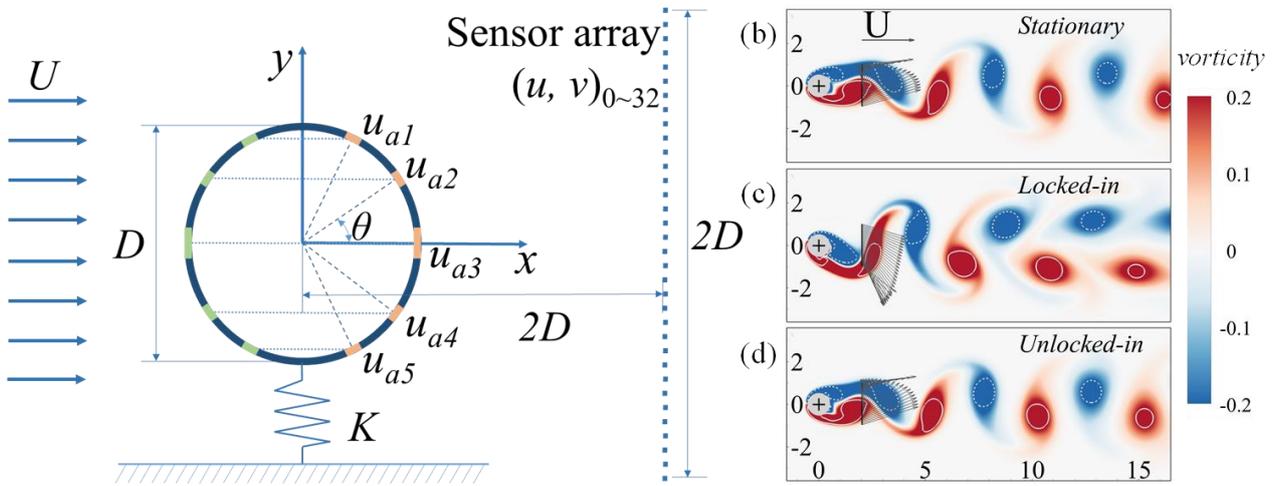

Figure 1. (a) Schematic of the AFC problem. Five pairs of WSLB actuators are adopted to control the wake of the cylinder, arranged symmetrically ($\theta = \pm80°$, $\pm40°$, and $0°$) about the cylinder's centerline. Each pair of actuators consists of a suction slot (green arc) at the windward side of the cylinder and a blowing slot (red arc) at the leeward side. (b) to (d) Wake patterns of stationary cylinder, locked-in VIV cylinder ($m^* = 2$, $U_R = 5$), and unlocked-in VIV cylinder ($m^* = 1.5$, $U_R = 3$), respectively, at the same instant when the cylinder reaches its equilibrium position in the upstroke and the transverse force it experiences changes its direction from downward to upward. The wakes are colored by vorticity. The vortices are identified using the $\lambda_{ci}$ criterion and enclosed by grey lines, where solid and dashed lines indicate positive and negative vortices, respectively. Velocity vectors measured by the sensor array are also shown to demonstrate strong hydrodynamic signature in the cylinder wakes.

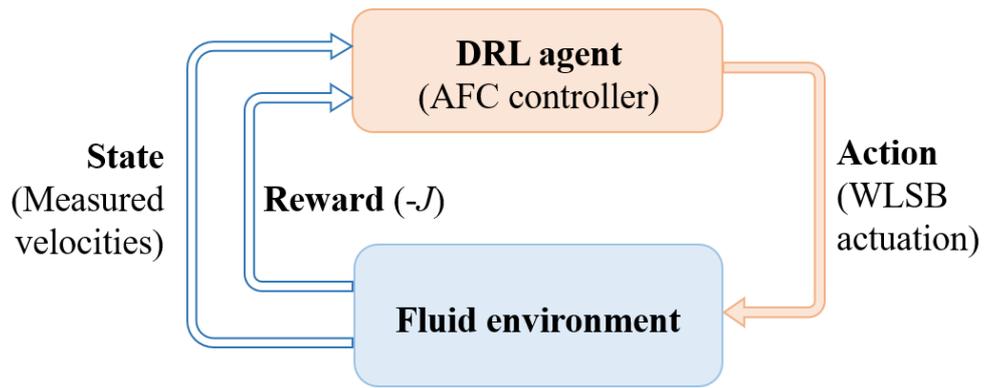

Figure 2. Schematic of the DRL loop used in the present study

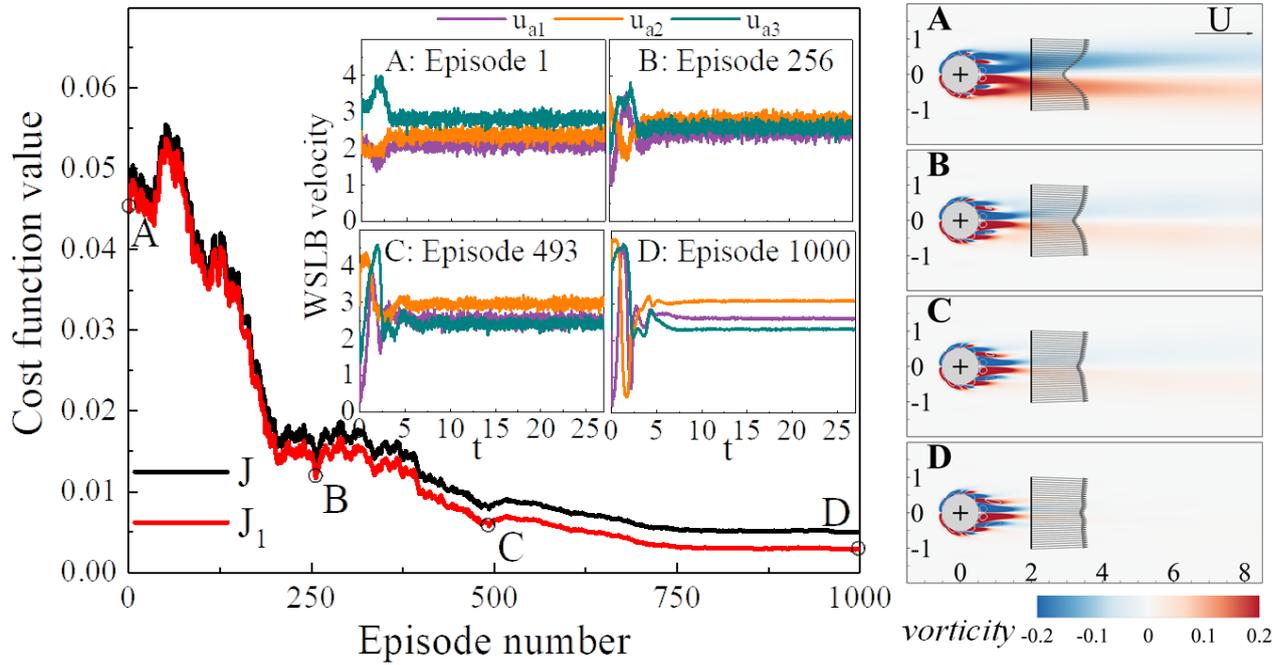

Figure 3. Left: Learning process represented by the variation of cost function values against episode number. The four insets show WSLB actuations generated by the DRL agent at different stages of the learning. Right: Instantaneous wake patterns and measured velocity profiles at the selected four stages. The progressive disappearance of velocity deficit well demonstrates the effectiveness of DRL trained wake control.

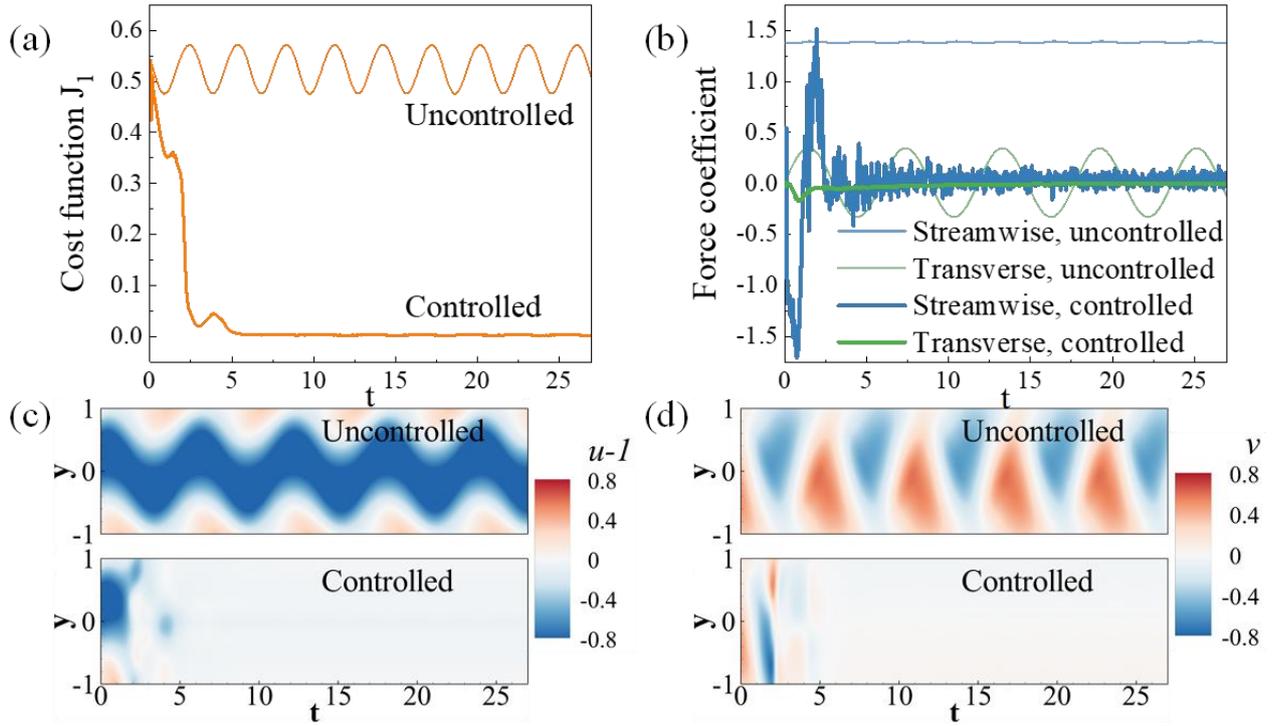

Figure 4. Evolution of $J_1$ value (a) and force coefficients (b) of uncontrolled and well-controlled cases for the stationary cylinder. The control strategy learnt in Episode 1000 (see Fig. 3) is adopted in the controlled case. The force coefficients were calculated by nondimensionalizing the streamwise and transverse components of hydrodynamic forces using $\rho U^2 D/2$. (c) and (d): Evolution of the corresponding net streamwise velocity, $u - 1$, and transverse velocity, $v$, measured by the sensor array, respectively. The control starts at $t = 0$.

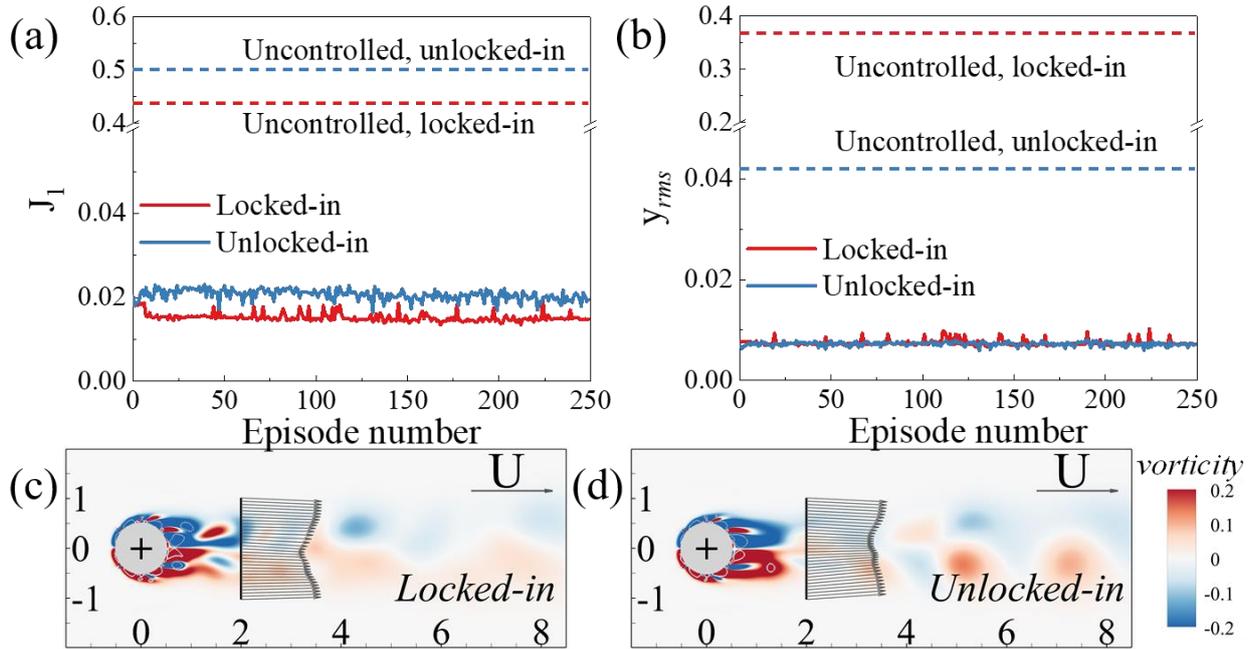

Figure 5. (a) Learning processes of DRL trained AFC on the locked-in and unlocked-in VIV cylinder. Since transfer learning is applied, both learning curves are nearly flat, indicating that the control strategy learnt from the stationary cylinder already works well for the VIV cylinder. (b) Variation of standard deviations of the cylinder's transverse displacement throughout the learning process, which are evaluated using the data in the second half of each episode. (c) and (d): Instantaneous wake patterns and measured velocity profiles for the well-controlled locked-in and unlocked-in VIV cylinder, respectively. In both cases, the control strategies learnt in Episode 250 are adopted.